# "The greatest Poet that has [n]ever existed"
# A Narrative Networks Analysis of the Poems of Ossian

Ralph Kenna, Pádraig MacCarron, Thierry Platini, Justin Tonra, and Joseph Yose

S URPRISING AS IT MAY SEEM, APPLICATIONS of statistical methods to physics were inspired by the social sciences, which in turn are linked to the humanities. So perhaps it is not as unlikely as it might first appear for a group of statistical physicists, applied mathematicians, and humanists to come together to investigate one of the subjects of Thomas Jefferson's poetic interests from a scientific point of view. And that is the nature of this article: a collaborative interdisciplinary statistical analysis of the works of a figure Jefferson described as a "rude bard of the North" and "the greatest Poet that has ever existed."

In 2012, a subset of this team embraced an increase in interdisciplinary methods to apply the new science of complex networks to longstanding and interesting questions in comparative mythology. Investigations of network structures embedded in epic narratives allowed universal properties to be identified and ancient texts to be compared to each other.[1] The approach—focusing on aggregate properties rather than individual characteristics, more commonly the focus of analysis in traditional studies—extracted statistical information about how characters are interconnected in such narratives.

These investigations motivated a new interdisciplinary direction of complexity science in digital humanities. Over the past number of years, it has evolved to address increasingly nuanced questions and in 2017, this team applied network theory to investi-gate the poems of Ossian,[2] one of Jefferson's lifelong interests. Other projects followed, such as a study of the Viking Age in Ireland, as recounted in medieval Irish texts.[3] Associated academic papers received significant media attention, kindling interest in old and ancient literature worldwide. The approach also inspired new challenges in mathematics, physics and even processes in industry, thereby illustrating how collaborations of this nature can be mutually beneficial and can capture the attention of a public, often ill-served by academic communication and dissemination. This article derives from these works, and from our consistent objective to help bridge the perceived gap between the natural sciences and the humanities.

---

[1] P. MacCarron and R. Kenna, "Universal Properties of Mythological Networks," *Europhysics Letters,* Vol. 99, 2012: 28002.

First we discuss the history of relationships between the sciences and the humanities and the tensions that exist in some quarters nowadays. Then we reflect upon the potentially unifying nature of a third culture: the new field of complex systems and statistical physics in particular. We discuss the origins of the poems of Ossian in the third section and Jefferson's interests in the next part, followed by our statistical approach in the fifth section. In the final section, we explore ideas for future research on these themes and discuss the potential of collaborative pursuits of human curiosity to overcome the two cultures dichotomy and embrace a scientific- and humanities-literate information age.

## Interdisciplinary Research
## Humanities and Statistical Physics

As cartographers have developed atlases of the physical world in which we live, and disciplinary maps of the human sphere of knowledge have also been drawn. Some of these place mathematics beside physics, and diametrically opposite to the humanities and social sciences,[4] but this view is not universal. Empirical research of the type essential to the natural sciences is neither a necessary nor a sufficient requirement to engage in mathematics, nor is it based on texts or culture as the humanities are. The question of where to place mathematics relative to physics and the humanities is connected with the age-old question of whether our reality is a mathematical structure or whether it is described by mathematics (i.e., a question of reality versus human perceptions of reality).[5] But such considerations are moot in the context of this article, for it is the same spirit of curiosity that motivates both scientists and humanists to pursue greater understandings of our world, be it physical or perceptual.

---

[2] J. Yose, R. Kenna, P. MacCarron, T. Platini, and J. Tonra, "A Networks-Science Investigation into the Epic Poems of Ossian," *Advances in Complex Systems,* Vol. 19, No. 2, 2012: 1650008.

[3] J. Yose, R. Kenna, M. MacCarron and P. MacCarron, "Network Analysis of the Viking Age in Ireland as Portrayed in *Cogadh Gaedhel re Gallaibh*," *Royal Society Open Science,* Vol. 5, 2018: 171024.

    This spirit of curiosity has arguably been impaired through increased metricization and monetization of research, researchers, and universities. Many scholars are pressured to succumb to forces that overlook curiosity and pursue metrics instead: to seek funding to meet targets with questionable relevance to the formerly avowed purpose of universities, thereby generating and imparting knowledge. As university bureaucrats seek to transform knowledge generation into profit generation at the behest of governments, the humanities are particularly vulnerable: the skills and knowledge they generate are perceived as "less tangible and effable than their science and engineering peers (and less readily yoked to the needs of the corporate world)."[6] Ironically, many government ministers and world leaders have humanities backgrounds. Yet some gov-

---

[4] E.g., K. Börner et al., "Design and Update of a Classification System: The UCSD Map of Science," *PLoS ONE,* Vol. 7, 2012: e39464.

[5] E.g., M. Tegmark, *"*The Mathematical Universe,*" Foundations of Physics*, 38 (2008) 101–50.

[6] A. Preston, "The War Against Humanities at Britain's Universities," *The Guardian*, 29 Mar. 2015, https://www.theguardian.com/education/2015/mar/29/war-against-humanities-at-britains-universities, accessed 25 Nov. 2018.

ernments view the humanities as outdated, unprofitable, and even elitist.

The tensions between science and the humanities are not new. At the 1959 Rede Lectures in Cambridge University, C. P. Snow lamented what he perceived as a "gulf of mutual incomprehension" between the two areas, instigating the famous "two cultures" debate.[7] Snow began his career as a physical chemist and went on to become a novelist, so he considered himself well placed to comment. He argued that the needs of humankind would be best served by science and technology which would deliver hope and prosperity for the future. But he saw this goal as being impeded by a divide between the two cultures and that this divide was worsened by educational structures, at least in the UK. Snow believed that although humanists and scientists both struggle to communicate with one another, most of the blame for this cultural chasm lay with the humanities. He famously used the term "natural Luddite" to describe his perception of how the scientist views the humanist.

At the Richmond Lecture at Downing College Cambridge in 1962, F.R. Leavis severely criticised Snow's arguments.[8] "Where is such 'hope' to be found," asked Leavis, "except in the lives lived by particular persons?" In a follow-on book, *A Second Look*, Snow discussed a scientific- and humanities-literate third culture involving various disciplines which address the impact of scientific progress on humans: architecture, economics, genetics, medicine, political science, psychology, sociology, and so on.[9] The third culture, he argued, can overcome the perceived divide through collaboration and cooperation, through mutual understanding, respect, and gain. Statistical physics and complexity science are well

---

[7] C.P. Snow, *The Two Cultures* (London: Cambridge University Press, 1959).

[8] F.R. Leavis, *Two Cultures: The significance of C.P. Snow, with Introduction by Stefan Collini* (Cambridge: Cambridge University Press, 2013).

[9] C.P. Snow, *The Two Cultures: And a Second Look: An Expanded Version of the Two Cultures and the Scientific Revolution* (Cambridge University Press, 1963), and *The Two Cultures with Introduction by S. Collini* (Cambridge: Cambridge University Press, 1998).

placed to contribute to the emergence of a third culture in academia.

Statistical physics is built on simplified bottom-up approaches—how the properties of large-scale systems emerge from interactions between basic components. Of course its scientists are also interested in the physics of microscopic objects such as atoms, but the statistical approach often grossly simplifies these by whittling them down to their bare essentials. In the famous Ising model of magnetism, for example, the quantum nature of the magnetic dipole moments of atomic spins is represented by simple binary states—up or down, quantified as +1 or -1.[10] Thanks to a remarkable phenomenon called universality, gross simplification at the micro level can deliver understandings of reality at the macro level provided the interactions between statistically large numbers of constituent parts are properly accounted for and the right questions are asked.

Although deeply embedded in fundamental physics, its combination with statistics opens up navigable pathways from statistical physics to other disciplines. Indeed, when it comes to the social sciences, it is rather a journey home for statistical physics; like the humanities, statistical physics has a cousin there. It was in the eighteenth century that certain regularities were observed in numbers of events such as births, marriages and deaths. People are very different at an individual level, so it came as a surprise that individuals were predictable in the aggregate. These curiosities partly motivated the subsequent development of statistical approaches to complex systems in physics. Notions such as these—aggregates, interactions and universality—remain the link between statistical physics and the social sciences. The power of modern computers to deal with many-body systems and the availability of big data explain why cross-disciplinary initiatives gathered pace. Indeed, in recent years interdisciplinary approaches have become increasingly popular as some physicists turn their attention from fundamental to complex problems. In the words of Stephen Hawking, we are now in the "century of complexity," moving on from fundamental laws that govern matter to how everything is connected to everything else.[11]

---

[10] E. Ising, "Beitrag zur Theorie des Ferromagnetismus," *Zeitschrift für Physik,* Vol. 31, 1925: 253–58.

In recent decades, the discipline of sociophysics has emerged and established itself as an academic area which offers new perspectives on many aspects of collective behaviour in society.[12] This is the spirit which led to a new, quantitative approach to studying old and ancient narratives, focusing on properties which emerge through interconnections between characters. Simplifying characteristics are recorded through sets of numbers, quantitative comparisons are possible, and these may yield meaningful answers to certain questions.

Any application of a new method across disciplinary boundaries may be expected to entail caveats and limitations and these should be obvious in the present instance too. As sociophysics is not a replacement for traditional approaches to sociology, neither can this approach capture the nuances of characters and their interactions. This approach cannot deliver information on events, emotions, meanings or other qualitative and holistic features which form the focus of close reading and forensic analysis. But it does deliver a bottom-up approach to the aggregate using only essential information at the micro-level to deliver macro-phenomena. In this sense, the new approach can complement existing techniques.

We can now turn to the main topic of this article: the Ossianic poems and the appeal they held for Jefferson. We examine these in the next two sections before proceeding to our statistical analysis. We refer throughout to the literary work simply as *Ossian* (italics) and to the character and purported author as Ossian (without italics).

---

[11] S.W. Hawking, in answer to a question: "Some say that while the twentieth century was the century of physics, we are now entering the century of biology. What do you think of this?" *San Jose Mercury News*, January 23, 2000.

[12] S. Galam, *Sociophysics: A Physicists Modeling of Psycho-Political Phenomena* (Berlin: Springer-Verlag, 2012).

## The Epic Poems of Ossian and their Origins

In 1760, James Macpherson published the first of a series of epic poems which he claimed to have translated into English from ancient Scottish-Gaelic sources dating from the "most remote antiquity."[13] The poems, he claimed, came from a third-century bard named Ossian and they quickly achieved international acclaim, finding prominent advocates such as Thomas Jefferson. Indeed, the poems have been described as among "the most important and influential works ever to have emerged" from Britain or Ireland,[14] and comparisons were drawn to major works of the epic tradition such as Homer's *Iliad* and *Odyssey*. However, doubts about their authenticity provoked one of the greatest literary controversies of all time, which reverberates in Ossianic studies to this day. The uncertainties about the works' authenticity centred on questions about the veracity of its claims to antiquity and its documentary origins, and on suspicions that the poems misappropriated material from Irish mythological sources. Although modern literary scholarship on *Ossian* has largely moved on from the original debate, the 250th anniversary of the poetry in the 1990s brought the Ossi-

---

[13] J. Macpherson, *Fragments of Ancient Poetry, Collected in the Highlands of Scotland, and Translated from the Galic or Erse Language* (Edinburgh: G. Hamilton and J. Balfour, 1760); J. Macpherson, *Fingal, an Ancient Epic Poem, in Six Books; Together with Several Other Poems, Composed by Ossian, the Son of Fingal. Translated from the Galic Language* (London: T. Beckett and P.A. DeHondt, 1762); J. Macpherson, *Temora, an Ancient Epic Poem, in Eight Books; Together with Several Other Poems, Composed by Ossian, the Son of Fingal. Translated from the Galic Language* (London: T. Beckett and P.A. DeHondt, 1763); J. Macpherson, *The Works of Ossian, the Son of Fingal, in Two Volumes. Translated from the Galic Language by James Macpherson. The Third Edition, To Which Is Subjoined a Critical Dissertation on the Poems of Ossian* (London: T. Becket and P.A. De Hondt, 1765); and J. Macpherson, *The Poems of Ossian, Translated by James Macpherson, Esq., in Two Volumes* (London: W. Strahan and T. Becket, 1773).

[14] H. Gaskill, ed., *The Poems of Ossian and Related Works* (Edinburgh: Edinburgh University Press, 1996).

anic controversy to the fore once again and the topic has undergone something of an academic renaissance since

The extent of *Ossian* and the comparisons made with epics from the Classics and Irish mythology prompt an invitation to investigate these works from a recently-developed networks-science point of view. We present a fresh investigation into Macpherson's famous works, based upon the new quantitative approach. In particular, we contrast the social-network structures that are unconsciously embedded in *Ossian* with works which have been identified as potentially playing influential roles in its composition. These are Homer's epics and Irish mythological texts: specifically, *Acallam na Senórach* (*Colloquy of the Ancients*), from the so-called Fenian Cycle in Irish mythology. Our aim is to determine whether or not the network structures that underlie the society depicted in Macpherson's work are similar to those of either corpus. We show that the Ossianic network is dissimilar to those of Homer but we reveal a remarkable structural similarity to the society underlying *Acallam.* This suggests a clear affinity between Macpherson's works and the Fenian Cycle from which he sought distance in the commentary which accompanies his texts.

In this section, we provide some context for the origins of the Ossianic controversy, explaining the circumstances around which the work appeared. We also give a short account of how it was received with both acclaim and condemnation. We further discuss the cultural significance, impact, and legacy of the work and why it remains an active area for academic interest over 250 years later. In the following section, we focus on Jefferson's interest in the texts, and related matters.

In the eighteenth century, the Highlands of Scotland were culturally different from the rest of Britain and continental Europe with distinct societal structure and customs. However, in 1746, defeat at the Battle of Culloden led to its assimilation into Great Britain and the suppression of native Gaelic culture and Jacobite clans. Against this backdrop of cultural depletion, Macpherson's first volume, *Fragments of Ancient Poetry, Collected in the Highlands of Scotland, and Translated from the Galic or Erse Language*, was published in 1760. By connecting its readers to a lost heroic age, scholars have read the work as motivated by a desire to mend some of the damage inflicted on the Scottish Highlands after the Jacobite Rising.[15]

The fragments were supposed to have originated from a blind third-century bard named Ossian, whose rhythmic tales of strife, war, and love evoked an ethereal atmosphere which greatly appealed to a receptive public. Indeed, aspects of this image of the Scottish Highlands persist to the present day.[16] Two further Ossianic volumes, *Fingal* and *Temora*, followed and in 1765, the corpus was collected in *The Works of Ossian, the Son of Fingal*. This included "A Critical Dissertation on the Poems of Ossian, the Son of Fingal," by Hugh Blair, Professor of Rhetoric at the University of Edinburgh. We base our network analysis in Section 5 on the text of the recent authoritative edition by Howard Gaskill.[17] Although a number of variants between various editions exist, these do not affect the character interactions from which our data derive.

The claims for an ancient and noble Scottish literary heritage were welcomed by Scottish scholars such as Blair, Adam Ferguson, and David Hume, as a catalyst for an increased sense of national identity when confronted by the cultural fragmentation of the Highland Clearances. Scholarly works such as Blair's *Critical Dissertation*, attempted to bolster the credibility and authenticity of the Ossianic poems. Thus, *Ossian* was of crucial importance in promoting Highland traditions and culture beyond the north of Scotland.

Impact beyond Scotland was enormous. Literary figures such as Blake, Byron, Coleridge, Goethe, Scott, and Wordsworth

---

[15] F. Stafford, "Introduction: The Ossianic poems of James Macpherson," in *The Poems of Ossian and Related Works*, H. Gaskill, (ed.) (Edinburgh: Edinburgh University Press, 1996).

[16] K. McNeil, *Scotland, Britain, Empire: Writing the Highlands, 1760–1860* (Columbus: The Ohio State University Press, 2007); P. Womack, *Improvement and Romance: Constructing the Myth of the Highlands* (London: Macmillan, 1989).

[17] H. Gaskill, ed., *The Poems of Ossian and Related Works*.

acknowledged *Ossian*'s influence. Composers such as Brahms, Mendelssohn and Schubert were inspired by the work, as were painters like Abildgaard, Gerard, Girodet, Ingres, Kauffmann, Krafft and Runge. The enthusiastic reactions of political leaders and public figures worldwide are recorded. Napoleon kept a copy in the portable library that he brought on military campaigns and Jefferson's famous statement that Ossian was "the greatest poet that has ever existed"[18] added to the chorus of approval for the poems.

*Ossian* was instrumental in rekindling interests in the traditional folklore, mythology, and poetry of various nations and through this conduit it inspired romantic nationalism. Appearing, as it did, to "vindicate a fallen nation"[19] it resonated in countries which had been subjected to conquest and which sought reminders of a glorious national past. Other nations were inspired to recover their own national literature. In 1764 Evan Evans's *Some Specimens of the Poetry of the Ancient Welsh Bards*[20] argued for a Welsh native poetry on a par with that of the Highlands. In 1775, Thomas Percy published *Reliques of Ancient English Poetry*[21] and Charlotte Brooke's *Reliques of Irish Poetry*[22] followed in 1789. The recep-tion of *Ossian* in wider Europe and its development in various contexts are explored in detail by Bär and Gaskill.[23]

---

[18] J.L. Golden and A.L. Golden, *Thomas Jefferson and the Rhetoric of Virtue* (Lanham, MD: Rowman and Littlefield, 2002).

[19] D. Quint, *Epic and Empire: Politics and Generic Form from Virgil to Milton* (Princeton: Princeton University Press, 1993).

[20] E. Evans, *Some Specimens of the Poetry of the Ancient Welsh Bards. Translated into English, with Explanatory Notes on the Historical Passages, and a Short Account of Men and Places mentioned by the Bards, in Order to Give the Curious Some Idea of the Tastes and Sentiments of Our Ancestors, and Their Manner of Writing* (London: R. and J. Dodsley, 1764).

[21] T. Percey, *Reliques of Ancient English Poetry*: Consisting of Old Heroic Ballads, Songs, and Other Pieces of our Earlier Poets (*Chiefly of the lyric kind.*) *Together with Some Few of Later Date* (London: J. Dodsley, 1775).

[22] C. Brooke, *Reliques of Irish Poetry*: *Consisting of Heroic Poems, Odes, Elegies, and Songs, Translated into English Verse, With Notes Explanatory and Historical, and the Originals in the Irish Character, to which is Subjoined an Irish Tale* (Dublin: J. Christie, 1816).

By the 1770s, Macpherson's supporters saw the similarities between *Ossian* and Homer's epics as having already been established. The affinity the poems shared with the *Iliad* was particularly clear.[24] Macpherson, with the probable assistance of Hugh Blair, provided comparative passages from the Classics in the footnotes to *Fingal*. Emphasising similarities between episodes in Homer and *Ossian* had an authenticating function, further boosted by Blair's 'Critical Dissertation.' The term "Homer of the North" (attributed to Madame de Staël) also captures the common association.[25]

The reaction in England was quite different, and was most notably expressed by Samuel Johnson. The emphasis on native literature in Macpherson's poems stood in stark contrast to Johnson's interests and experience of the Classics. He famously identified a "strong temptation to deceit" in the Ossianic poems,[26] and in 1773, he travelled to the western islands of Scotland to investigate their purported origins. He believed that ancient Scottish vernacular manuscripts did not exist and that poems of the scale of *Ossian* could not have been passed intact orally through so many generations.

However, Johnson's reference to Gaelic as "the rude speech of barbarous people" betrayed widespread English hostility towards Scottish culture and undermined the credibility of his arguments. British scholars and administrators of the Imperial Era tended to align their attitudes towards the Classics and to view themselves as inheritors of the torch of civilization, handed down from ancient

---

[23] H. Gaskill, ed., *The Reception of Ossian in Europe* (London: Thoemmes, 2004).

[24] H. Gaskill, "The Homer of the North," *Interfaces,* Vol. 27, 2007–2008: 13–24.

[25] S. Mitchell, "Macpherson, Ossian, and Homer's Iliad," *Ossian and National Epic*, ed. G. Bär and H. Gaskill (Frankfurt am Main: Peter Lang, 2012.

[26] S. Johnson, *A Journey to the Western Islands of Scotland* (London: A. Strahan & T. Cadell, 1775).

Greece and Rome. They saw Homer as the epitome of this culture and Greco-Roman civilization. In comparison, Scottish culture was viewed as inferior. They justified conquest and colonization of Scotland as replacing an expendable culture with something superior. If an ancient epic tradition emanated from the Highlands, this would have challenged these attitudes. Besides a contest between Classicism and Romanticism, the debate therefore was linked to the clash between two emerging national identities within Great Britain.

Reaction to *Ossian* in Ireland was even more pronounced, but for different reasons. It was obvious to Irish antiquarians that the poems were drawn from tales of the Fenian Cycle. Thinly disguised characters, places, and situations from the Irish epic tradition were readily identified. In 1763, Sylvester O'Halloran denounced Macpherson's deceptive attempt to misappropriate Ireland's culture and its Gaelic heroes for Scotland.[27] The antiquarian and Gaelic scholar Charles O'Conor further championed the Irish cause.[28] For example, the character Ossian—Macpherson's "illiterate Bard of an illiterate Age"—was identified as Oisín, a warrior-poet of the Fenian Cycle in Irish mythology. Ossian's father, Fingal, who was a third-century Scottish king in Macpherson's texts, was identified as Fionn mac Cumhaill, leader of the Fianna Éireann, a heroic warrior band. Macpherson's Cuchullin, "general or Chief of the Irish tribes," is Cú Chulainn, the hero of an entirely different Cycle in Irish mythology. The Irish were further affronted by how Macpherson inverted the ancient relationship between the two countries, reversing the migration of populations and folklore.

---

[27] S. O'Halloran, "The Poems of Ossine, The Son of Fionne MacComhal, Re-claimed," *Dublin Mag.* (1763) 21–23; reprinted in Vol. III of D. Moore, *Ossian and Ossianism* (London: Taylor and Francis, 2004), 87–89.

[28] C. O'Connor, *Dissertations on the History of Ireland*; *To Which is Subjoined, a Dissertation on the Irish Colonies Established in Britain with Some Remarks* (Dublin: G. Faulkner, 1766).

## Ossian and Jefferson's Interests

When Thomas Jefferson's friend and brother-in-law, Dabney Carr, died at the age of 29 in 1773, Jefferson set about composing an inscription for his tombstone.[29] Before adapting an inscription from Scottish poet David Mallet's "The Excursion," Jefferson drafted a dedication inspired by another Scottish poem:

> This stone shall rise with all it's [*sic*] moss and speak to other years "here lies gentle Carr within the dark and narrow house where no morning comes with her halfopening pages." When thou, O stone, shalt fail and the mountain stream roll quite away! Then shall the traveller come, and bend here perhaps in rest. When the darkened moon is rolled over his head, the shadowy form may come, and, mixing with his dreams, remind him who is here.

Jefferson adapted this passage from Book II of *Temora*, the third publication of the Ossianic corpus, and supplemented it with characteristic Ossianic phrases that he recorded in his *Literary Commonplace Book*.[30] The sentiments and diction of the original Ossianic passage captivated Jefferson and have perplexed some Jeffersonian scholars who found in Ossian's sentimental appeal the antithesis of Jefferson's rational and analytic intellect. How could such a work—on the one hand a repository of sentiment and sensibility, and on the other, the subject of the eighteenth century's greatest literary scandal of authenticity—be for Jefferson "the source of daily and exalted pleasures"?[31] How could he believe this poet, whose very existence was apparently so uncertain, "the greatest Poet that has ever existed"?[32]

---

[29] Susan Manning, "Why Does It Matter than Ossian Was Thomas Jefferson"s Favorite Poet?" *Symbiosis*, Vol. 1, No. 2, 1997: 230; Jack McLaughlin, "Jefferson, Poe, and Ossian," *Eighteenth-Century Studies*, Vol. 26, No. 4, 1993: 630–31; and Thomas Jefferson, *Jefferson's Literary Commonplace Book,* ed. D.L. Wilson (Princeton: Princeton University Press, 2016), 9 and 172.

[30] Thomas Jefferson, *Jefferson's Literary Commonplace Book*, 144.

*Ossian*'s position within Jefferson's intellectual and aesthetic consciousness is undoubtedly worthy of remark. While some have argued that Jefferson's Ossianic fascination has its origin in his youth and should not be afforded excessive attention,[33] evidence of its enduring influence well into the period of his public life suggests that it should neither be ignored nor dismissed. Scottish influences are present in Jefferson's elementary and further education.[34] From the age of nine, he was educated by Rev. William Douglas, graduate of Glasgow and Edinburgh Universities and tutor for the beginnings of Jefferson's classical schooling. At William and Mary, he was taught by Dr William Small, a Scottish devotee of the principles of Glaswegian moral philosopher, Frances Hutcheson. The influence of classical literature and the works of the Scottish Enlightenment on the young Jefferson is evident in his *Literary Commonplace Book*, where David Hume sits alongside Homer, Virgil, and Cicero.

The *Literary Commonplace Book* is the document that contains Jefferson's most concentrated engagement with *Ossian*: fourteen of the book's 407 entries are taken from Ossian's works. Likely originating as a scholarly exercise around the age of fifteen,[35] the *Literary Commonplace Book* was maintained by Jefferson until around the age of 30. The majority of its entries are poetic, and that shows his profound interest in poetry while young. That said, the demands of public life gradually robbed Jefferson of his enthusi-asm for literature,[36] as the conclusion of the *Literary Commonplace Book* in the mid-1770s illustrates. However, he did remark that his acquaintance with the darker aspects of life, encountered while practicing law, were alleviated by reading poetry.[37]

---

[31] Gilbert Chinard, "Jefferson and Ossian," *Modern Language Notes*, Vol. 38, No. 4, 1923: 202.
[32] Gilbert Chinard, "Jefferson and Ossian," 202.
[33] Thomas Jefferson, *Jefferson's Literary Commonplace Book*, 3.
[34] Thomas Jefferson, *Jefferson's Literary Commonplace Book*, 4. Paul J. Degategno, "'The source of daily and exalted pleasure': Jefferson Reads the Poems of Ossian," *Ossian Revisited*, ed. Howard Gaskill (Edinburgh University Press, 1991), 98.
[35] Thomas Jefferson, *Jefferson's Literary Commonplace Book*, 4.

We also have ample evidence of Jefferson's continued use of his *Literary Commonplace Book* after he ceased making regular contributions in 1773. Beyond its role in drafting inscriptions for Dabney Carr's tombstone, its texts provided the fodder for inscriptions for himself and his sister, Jane.[38] After being bound in Philadelphia in the early 1780s, the *Literary Commonplace Book* accompanied Jefferson to Paris in 1784 and influenced his 1786 essay, "Thoughts on English Prosody," composed during his ministry. Excerpts from passages recorded in the book appear in Jefferson's letters, and are used as late as 1816, when he sent passages from Horace and Cicero as samples of his handwriting to Amos J. Cook, a Maine schoolmaster.[39] Though his interest in literature may have waned as his commitment to public service grew, abundant evidence affirms the persistent presence of the *Literary Commonplace Book* in Jefferson's consciousness.

Ossian's place alongside classical authors in this book closely echoes Macpherson's strategy in presenting parallels for his Scottish epic in the literature of Ancient Greece and Rome. By thus comparing Ossian's words with canonical models, Macpherson sought esteem by association, while circumventing more obvious similarities between Ossian and the tales of the Irish Fenian cycle. Jefferson's interest in comparative narrative and mythology differed from Macpherson's strategic purposes, however. Rather, the entries in the *Literary Commonplace Book* suggest that he was drawn to recording examples of similar themes, images, and narratives from across literary history. His attention, Douglas Wilson argues, was drawn to "allusive relationships and the ways in which literary works anticipate and reflect one another."[40] Far from an aberration or exception within his intellectual life, this intertextuality *avant la lettre* enabled Jefferson to examine the aesthetic expression of enduring universal ideas found elsewhere amid his moral and philosophical interests.

---

[36] Thomas Jefferson, *Jefferson's Literary Commonplace Book*, 4.
[37] Thomas Jefferson, *Jefferson's Literary Commonplace Book*, 8.
[38] Thomas Jefferson, *Jefferson's Literary Commonplace Book*, 9 and 130.
[39] Thomas Jefferson, *Jefferson's Literary Commonplace Book*, 10.

The most common point in analyses of Ossian's influence on Jefferson is in connection to the narrative of Logan, the defeated Native American leader, described in *Notes on the State of Virginia*. Jefferson recounts the "Lament of Logan" in terms reminiscent of the rhetoric and diction of Ossian. After signing a peace treaty to end his people's conflict with Lord Dunmore's colonial troops in 1774, Logan delivered a speech detailing the woes of a leader contemplating the extinction of his race. It concluded:

> There runs not a drop of my blood in the veins of any living creature. This called on me for revenge. I have sought it: I have killed many: I have fully glutted my vengeance: for my country I rejoice at the beams of peace.[41] But do not harbour a thought that mine if the joy of fear. Logan never felt fear. He will not turn on his heel to save his life. Who is there to mourn for Logan?—Not one.[42]

By valorising the speech and its orator, Jefferson constructed, according to Amanda Louise Johnson, an "imaginary Indian" whose nobility balanced the melancholy of his defeat. Jefferson's noble fiction, Johnson continues, was a convenient symbol intended to efface the fate of the living Native Americans whose presence was perceived by settlers as a hindrance to the westward expansion of the Anglo-American republic. To transfer the Ossianic aesthetic to an American setting, Jefferson compared Logan's speech to those of Demosthenes and Cicero, following Macpher-

---

[40] Thomas Jefferson, *Jefferson's Literary Commonplace Book*, 11.

[41] This phrase has particular Ossianic resonance, as the expression "beam[s] of" occurs 136 times in the *Ossian* corpus.

[42] Thomas Jefferson, *Notes on the State of Virginia* (London: John Stockdale, 1787), 106.

son's model of ennobling Ossian and his highland kin by comparison of the bard's words to Homer's.[43]

Though Jefferson demonstrated little interest in debates about the authenticity of the *Ossian* poems, his own account of Logan's speech was subjected to similar accusations of fraudulence and forgery: to the extent to which, as James and Alan Golden describe,[44] American soldiers involved in the campaign against Logan and his family sought to defend themselves by rebutting the accuracy of his lament and the nobility it implied. In his counter-defence, Jefferson published corroborating evidence from witnesses in an appendix to a later edition of the *Notes on the State of Virginia*, and resorted to the type of rhetorical ambiguity employed by Macpherson: "whether Logan's or mine, it still would have been American."[45] The characteristics of *Ossian* which appealed to Jefferson, Johnson argues, are bound in its mythic removal from reality and in its repetitive and self-renewing structure. The artifice of the narrative techniques provided a model for Jefferson's objectives to "assert the Republic as a civilisation *de novo*, while still maintaining its inherited British cultural and legal institutions, and establish the Anglo-American's organic connection to American soil in contradiction to their immigrant past."[46] The recursive manner in which the bard, Ossian, narrates the decline of his civilisation suggested a rhetorical structure to Jefferson, wherein the civilizational breaches and destructions which enabled the new republic could be mythologised into a narrative of renewal. Much more than youthful infatuation, Johnson concludes, Jefferson's enduring interest in Ossian lay in its provision of a rhetorical mod-

---

[43] Amanda Louise Johnson, "Thomas Jefferson's Ossianic Romance," *Studies in Eighteenth-Century Culture*, Vol. 45, No. 1, 2016: 28.

[44] J. L. Golden and A. L. Golden, *Thomas Jefferson and the Rhetoric of Virtue* (Lanham, MD: Rowman and Littlefield, 2002), 356–72.

[45] Amanda Louise Johnson, "Thomas Jefferson's Ossianic Romance," 29, and Susan Manning, "Why Does It Matter that Ossian Was Thomas Jefferson"s Favorite Poet?" 229.

[46] Amanda Louise Johnson, "Thomas Jefferson's Ossianic Romance," 24.

el for promoting the national and imperial goals of the young American republic.[47]

While the repetition and recursion of *Ossian* is central to Johnson's reading of Jefferson's interest, Susan Manning focuses on these and other features—dislocation, ambiguity, and nebulousness—which critics of the work have identified as evidence of its fraudulence. Manning argues that these are the very characteristics which appealed to Jefferson and which proved influential as "a means of distancing and harmonizing sentiments with an uncomfortably discordant and specific charge of grief and guilt."[48] Johnson and Manning agree on Logan's value to Jefferson and the republican project residing in his symbolism of "virtuous eloquence,"[49] but Manning argues for the additional significance of translation in presenting both Ossian and Logan. As both as translated to be rendered in the language of "polite" and "civilised" discourse, translation marks and masks the loss of the original Gaelic and Iroquois languages[50]: "There is no 'original': the text is itself a metaphor for that loss."[51] Manning and Johnson agree that Jefferson's rendering of Logan's speech is an articulation of extinction within the framework of the destroyer's discourse, with Manning adding that the subsequent inclusion of Logan's speech in editions of McGuffey's Readers was one token of the way in which "enlightened Americans … compensated for their destruction of native culture by simultaneously reincorporating it within an idiom of sentimental conservatism."[52] Manning's argument departs from Johnson's in its presentation of Jefferson's Ossianic tombstone inscription as further evidence of the work's distancing effects: this time in the event of grief. Many scholars have noted Jefferson's use of an Ossianic excerpt in a proposed inscription for Dabney Carr's tombstone, but Manning alone places the significance of this gesture in the same context as Logan's speech. Jack McLaughlin also discusses 'Logan's Lament' in connection to his analysis of *Ossian*'s influence on Jefferson. However, his analysis is less extensive than that of Manning and Johnson, focusing instead on the stylistic and syntactic similarities in the two texts, and concluding that the "sentimental primitivism"[53] of Ossianic poetry provided him with an ideal template for his romanticised depiction of the Native American leader.

---

[47] Amanda Louise Johnson, "Thomas Jefferson's Ossianic Romance," 31.

[48] Susan Manning, "Why Does It Matter that Ossian Was Thomas Jefferson"s Favorite Poet?" 220.

[49] Susan Manning, "Why Does It Matter that Ossian Was Thomas Jefferson"s Favorite Poet?" 221.

[50] Susan Manning, "Why Does It Matter that Ossian Was Thomas Jefferson"s Favorite Poet?" 223.

[51] Susan Manning, "Why Does It Matter that Ossian Was Thomas Jefferson"s Favorite Poet?" 225.

[52] Susan Manning, "Why Does It Matter that Ossian Was Thomas Jefferson"s Favorite Poet?" 225.

If commentators identify Jefferson's interest in *Ossian* as a sentimental anomaly in an intellectual life otherwise characterised by rationality, McLaughlin locates the latter method in Jefferson's pursuit of Ossianic manuscripts. McLaughlin asserts that once Jefferson had decided on the greatness of "this rude bard of the North,"[54] his desire to see original manuscripts and understand the nature of the apparent greatness was indicative of the regular mode of Jeffersonian inquiry. Gilbert Chinard first detailed the correspondence initiated by Jefferson in 1773 in an attempt to acquire copies of the Gaelic manuscripts for personal study. Writing to Macpherson's brother, Charles, Jefferson claimed that "merely for the pleasure of reading his works I am desirous of learning the language in which he sung and of possessing his songs in their original form."[55] Those acquainted with the Ossian controversy will be unsurprised to learn that Jefferson's request was not satisfied. Macpherson replied to his brother that he could not "give a copy of the Gaëlic manuscripts, without any decency, out of my hands,"[56] and Charles relayed the news to Jefferson, thereby offering, by way of compensation, a Gaelic New Testament to satisfy his interest in the language.

---

[53] Jack McLaughlin, "Jefferson, Poe, and Ossian," 631.
[54] Jack McLaughlin, "Jefferson, Poe, and Ossian," 632.
[55] Gilbert Chinard, "Jefferson and Ossian," 202.
[56] Gilbert Chinard, "Jefferson and Ossian," 202.

Though Jefferson's desire to see the manuscripts echoed the terms of Samuel Johnson's critique of Ossianism,[57] he betrayed no trace of interest in querying the authenticity of the work. Indeed, a strikingly modern sophistication reverberates from his suggestion, as Amanda Louise Johnson relates, "that something belated and supposedly manufactured could have the same level of aesthetic value as something that was 'authentic' and original."[58] Jefferson's letter to Charles Macpherson, composed with a level of care and revision unique to Jefferson[59] testifies at once to the gravity of his inquiry and to his determination to appease *Ossian*'s translator: "He wants Macpherson to understand that he has been the best kind of reader, once who has attuned himself to the richness of *Ossian*'s language."[60] The correspondence is deemed "remarkable" by Jefferson scholars because it bears witness to "the extraordinary place in [his] literary pantheon occupied by *Ossian*."[61] Many scholars mention the episode but do not dwell on its significance beyond the unusual combination of Jeffersonian ardour and sagacity that it illustrates. Nonetheless, the exchange is significant for Ossianic studies because of the evidence it presents about the inscrutable *Ossian* manuscripts. Howard Gaskill reads James Macpherson's response to Charles and urges us to see beyond the apparent "embarrassed evasion and lame excuse of one whose cupboard is bare."[62] In Macpherson's reluctance to allow a copy, and especially in his reference to "*my* manner and *my* spelling," Gaskill sees an implicit acknowledgement that the "originals" comprised a Gaelic text that had been prepared by Macpherson himself. In the long and contentious debate about Ossianic manu-scripts and originality, Gaskill's argument is important and unduly overlooked.

---

[57] Samuel Johnson, *A Journey to the Western Islands of Scotland* (London: A. Strahan & T. Cadell, 1775), 267–77.
[58] Amanda Louise Johnson, "Thomas Jefferson's Ossianic Romance," 20.
[59] Gilbert Chinard, "Jefferson and Ossian," 201–2.
[60] Paul J. Degategno, "The Source of Daily and Exalted Pleasure,'" 99.
[61] Thomas Jefferson, *Jefferson's Literary Commonplace Book*, 13.
[62] Howard Gaskill, "What Did James Macpherson Really Leave on Display at his Publisher's Shop in 1762?" *Scottish Gaelic Studies*, Vol. 16, 1990: 72.

The very essence and nature of Jefferson's interest in *Ossian* has provoked repeated inquiry because it seems out of character. A degree of incredulity is evident in some of those inquiries: in the title of Manning's article, in McLaughlin's pairing of Ossian with Edgar Allan Poe, another unlikely recipient of Jefferson's attention. What may have begun in youth as an aesthetic infatuation possessed a durability that extended well into maturity, as the consonance of *Ossian* with Jefferson's broader thought revealed itself. This "ratio of sensibility to intellect"[63] is present in the anecdote from Jefferson's reception of the Marquis de Chastellux, a member of the French Academy and a general in Rochambeau's army,[64] at Monticello in 1782:

> I recall with pleasure, that as we were conversing one evening over a 'bowl of punch,' after Mrs. Jefferson had retired, we happened to speak of the poetry of Ossian. It was a spark of electricity which passed rapidly from one to the other; we recalled the passages of those sublime poems which had particularly struck us, and we recited them for the benefit of my traveling companions, who fortunately knew English well and could appreciate them, even though they had never read the poems. Soon the book was called for, to share in our 'toasts'; it was brought forth and placed beside the bowl of punch. And, before we realized it, book and bowl had carried us far into the night.[65]

Amid the etiquette of diplomacy and the propriety of public life, Jefferson responded with undisguised enthusiasm to the mention of the poems of Ossian. The shared admiration of *Ossian* was a catalyst for forging a bond of friendship between de Chastellux and his Monticello host, and a reminder of the unique place that this poetic work held, undiminished, throughout Jefferson's life.

---

[63] Jack McLaughlin, "Jefferson, Poe, and Ossian," 634.
[64] Paul J. Degategno, "The Source of Daily and Exalted Pleasure," 100, and Atcheson L. Hench, "Jefferson and Ossian," *Modern Language Notes*, Vol. 43, No. 8, 1928: 537.
[65] Thomas Jefferson, *Jefferson's Literary Commonplace Book*, 172–73.

## Networks and Narratives

Thus we come to the question that escaped the main focus of Jefferson's interest in Ossian: Was he a "Homer of the North," as he was dubbed by de Staël, or an "imposture" as per O'Halloran? Questions like this rather reflect the polar debate of the time than the direction of scholarship since. Although modern interests are more nuanced than a binary battle, it is nonetheless interesting to peruse old questions in a modern light given the new tools at our disposal. We wish to determine whether the new narrative-network approach can deliver results consistent with extant knowledge. That it does can inspire and encourage new questions, including ones closer to Jefferson's interests. For these reasons, we investigate the structural foundations of *Ossian* from a network-science perspective.

As discussed, there were two main comparisons to *Ossian* from the outset: the Homeric epics and the Fenian Cycle of Irish mythology. Previous research focuses mainly on individual characters in the narratives; our approach in this section is complementary in the sense that it addresses the entirety of the social networks embedded in them. Before presenting our results, we provide further context for the comparative texts.

The *Iliad* dates from the final year of the Trojan War (the eighth century B.C.) and tells of a dispute between Agamemnon, leader of the Greeks, and their hero Achilles. The *Odyssey* describes the journey home of Odysseus to his wife Penelope after the fall of Troy. The Fenian Cycle of Irish mythology recounts the adventures of Fionn mac Cumhaill and the warrior band, the Fianna. *Acallam na Senórach* (*Colloquy of the Ancients* or *Tales of the Elders*) is its most important source and describes how Fionn and his warriors meet the recently arrived Saint Patrick[66] and how the stories came to written form.

We use recent versions and translations: Gaskill's version of *Ossian*; Rieu's translation of the *Iliad*[67]; Shewring's translation for the *Odyssey*,[68] and a recent translation of *Acallam na Senórach*.[69] We also examine *The Fianna*, Part II of *Lady Gregory's Complete Irish Mythology*.[70] The latter is certainly derived from the Fenian Cycle and therefore any attempt at structure comparison of two texts should place it close to *Acallam na Senórach*. We will see that our method succeeds in passing this test. Other versions or translations may deliver differences in minor characters and relationships but the networks are expected to be robust to such variations because of the statistically large numbers of characters from which the data derive.

Our objective is to compare the network structures in *Ossian*, the *Iliad*, *Odyssey*, *Acallam na Senórach* and Lady Gregory's text. Many quantitative measures of network properties have been developed since network science emerged more than twenty years ago.[71] We have measured many network properties, and the results have been published by Yose et al.[72] but the essential information for our purposes is contained primarily in one feature: the degree distribution. To understand this, we must firstly examine how complex networks compare to simpler structures.

---

[66] For an accurate date, see D. McCarthy, "Analysing and Restoring the Chronology of the Irish Annals."

[67] *Homer, The Iliad,* trans. E.V. Rieu (London: Penguin Classics, 2003).

[68] *Homer, The Odyssey,* trans. W. Shewring (Oxford: Oxford University Press, 1980).

[69] A. Dooley and H. Roe, *Tales of the Elders of Ireland: A New Translation of* Acallam na Senórach (Oxford: Oxford University Press, 1999).

[70] A. Gregory, *Gods and Fighting Men: The Story of the Tuatha De Danaan and of the Fianna of Ireland*, trans. Lady Augusta Gregory (London: J. Murray, 1904), and reproduced in Lady Gregory's *Complete Irish Mythology* (Vacaville, CA: Bounty Books, 2004).

[71] D.J. Watts and S.H. Strogatz, "Collective dynamics of "small-world" networks," *Nature* 393 (1998), 440–42.

[72] J. Yose, R. Kenna, P. MacCarron, T. Platini, and J. Tonra, "A Networks-Science Investigation into the Epic Poems of Ossian."

The simplest possible networks include regular grids and random graphs (see Fig.1). The middle panel contains an example of network nodes randomly connected by links. Although they appear very different at first sight, the left and middle networks are actually similar in that they are both essentially uniform; besides random fluctuations in the middle graph, they are both homogeneous. The right network is quite different; few of its nodes have a large number of links and most nodes have a few. This is the type of inhomogeneity that occurs in society, for example, where a few famous people have large numbers of acquaintances, and most people have fewer.

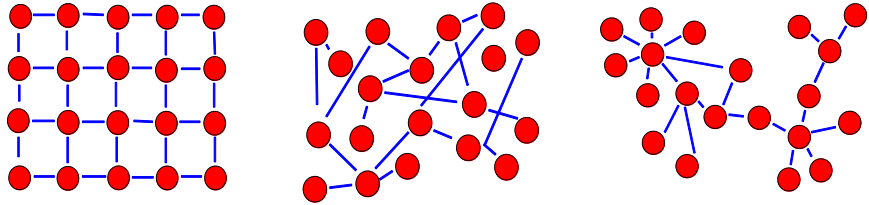

**Figure 1.** Examples of a regular lattice (left), a random graph (middle), and a complex network (right).

To construct a narrative network, we first carefully read the text (a number of times), taking note of individual characters and meticulously recording the interactions between them. We do not require details such as the personalities or complexities of individual characters. Instead we strip them down to their essentials and represent them by zero-dimensional nodes or points, because the set of interrelationships between characters, and not individual characters, primarily concerns us. Although shelving interesting information about characters that surely appealed to Thomas Jefferson might at first sight give the impression that we have stepped over the mark, note that every one of us does it every time we partake in democracy. To make collective decisions on highly complex and nuanced questions, we often ask the simplest possible questions whose answers are "yay" or "nay."[73] There are many other examples of binarizing the continuous, and the practice is not limited to network theory. In a recent trial of the effects of adrenaline on out-of-hospital cardiac arrest patients, for example, neurologic outcomes are categorised as favorable or not, binarizing a scale that ranges from the absence of symptoms to death.[74] Another example is that of a study of over 6,500 trees in a pistachio orchard. Yields of individual trees are binarized as above or below average, a simplification which was the basis for correlations between neighbouring trees to be explained by Ising-type interactions, discussed above.[75]

The first of these three examples involves a network: the network of society. Although the question pertains to individual preferences only, preferences are correlated across societal links. The second example also involves people but no network is involved: unlike voting preferences, one person's cardiac arrest is not likely to affect another's. The third example is a network, albeit not obviously. Subterranean fungal links or chemical responses to animal foraging are possible causes for correlations between plant yields.[76]

Besides illustrating the unifying nature of interdisciplinarity, these examples demonstrate the ubiquity of binary statistics, and its power. The robustness of the method depends on the presence of a large enough population to reduce the significance of individual subtleties. We found 325 characters (nodes) in Gaskill's version of the *Ossianic* texts: enough to perform a meaningful analysis. We know from statistical concepts of sample size that once our networks are large enough, small differences between versions or editions of a work will not affect the outcomes of our study. To set up the network, a reader has to decide on the nature of their relationship: if pairs of characters are related; if they speak directly to each another, or speak about one another; of if they are physically are co-present and it is clear that they know each other. In these cases the nodes that represent them are joined by a link (also called an "edge"). If not, no link/edge is inserted. The nodes and the edges together form the *network*. Of course, a degree of subjectivity is involved in determining these links because they are sometimes implicit rather than explicit. We gather data using unautomated methods, but even close reading can sometimes fail to identify the nature of individual relationships with absolute certainty. Studies on robustness suggest that a large sample size delivers sufficient statistical power and small deviations in lesser prominent individual nodes and links are compensated by the scale of the study.[77] Attempts have been made to construct narrative networks using Natural Language Processing (NLP), a subfield of Artificial Intelligence (AI) but these have not so far been successful.[78]

---

[73] K. Wiesner et al., *European Journal of Physics,* Vol. 40, 2019, 014002.

[74] G.D. Perkins et al, "A Randomized Trial of Epinephrine in Out-of-Hospital Cardiac Arrest," *New England Journal of Medicine,* 2018: 379 and 711–21.

[75] Andrew E. Noble, Todd S. Rosenstock, Patrick H. Brown, Jonathan Machta, and Alan Hastings, "Spatial Patterns of Tree Yield Explained by Endogenous Forces Through a Correspondence Between the Ising Model and Ecology," *Proceedings of the National Academy of Sciences,* Vol. 115, 2018: 1825–30.

[76] We are grateful to David Arundel for pointing out this type of "willow talk" in African farming communities. See also, e.g., P. Wohlleben, *The Hidden Life of Trees* (London: William Collins, 2016).

The entire Ossianic network is displayed in Figure 2. Our evaluation of two comparisons in Section 4 above (namely Jefferson's comparison of Logan's speech with Demosthenes's and Cicero's on the one hand, and Macpherson's comparison of Ossian's poetry with Homer's on the other) represents a type of complexity that involves zooming in on the details of individual characteristics. In examining a full network, we address a very different type of com-plexity: how the edges are distributed between the nodes of the network. The simplicity of the latter type of complexity is that it utilises the totality of the interrelationships between characters in the texts. The two approaches are not in competition with each other but complementary. The real test of the new methodology is whether it delivers the same outcomes in circumstances where outcomes are known from the older one. And we demonstrate, in this paper at least, that it does.

---

[77] Investigations of network robustness were pioneered in the context of public transport networks in B. Berche, C. von Ferber, T. Holovatch, and Y. Holovatch, "Resilience of Public Transport Networks Against Attacks," *European Physics Journal B,* Vol. 71, 2009: 125–37.

[78] Marcello Trovati and James Brady, "Towards an Automated Approach to Extract and Compare Fictional Networks: An Initial Evaluation," 2014: 58, https://www.researchgate.net/publication/272745341_Towards_an_Automated_Approach_to_Extract_and_Compare_Fictional_Networks_An_Initial_Evaluation, accessed 10 Feb. 2019.

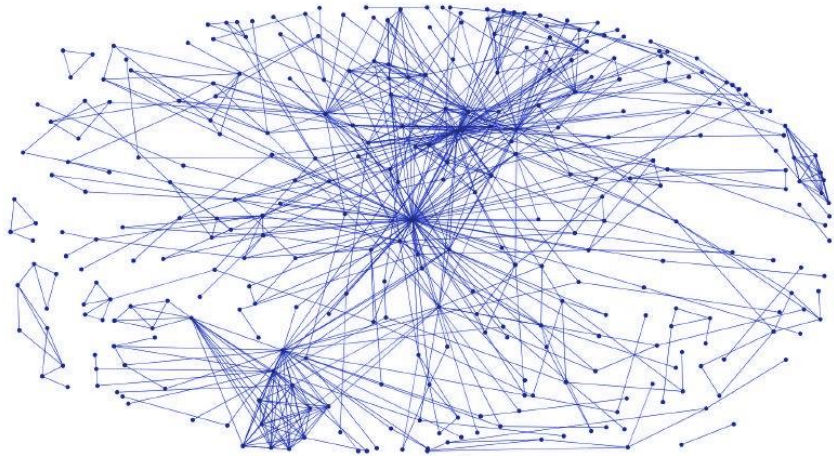

**Figure 2.** The entire network of 748 relationships between 325 characters depicted in Gaskill's version of Macpherson's Ossian.

To understand how the edges are distributed across the network, we offer Figure 3 as an example of a simple complex network. It is simple because it is small and easy to understand and complex as its topology is non-trivial. We usually denote the number of nodes by *N* and the number of links or edges by *M*. So, for example, in Fig.2 there are N=8 nodes and M=11 links between them.

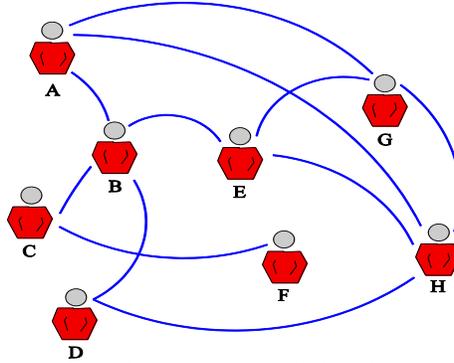

**Figure 3.** Networks comprise both nodes and links which represent relationships between them. In network theory, we are interested in statistically capturing how these relationships are distributed.

One of the most basic but essential network measures is the so-called degree of each node. This is the number of links that emanate from that node. We denote the degree of an individual node generically by *k* and if we identify the particular node in question we insert a subscript to label it. For example, in Figure 3, character A is linked with three other nodes (namely B, G and H). We say the degree *k* of node A is three or $k_A=3$. Different nodes have different numbers of edges attached to them and the average value of the degrees of all the nodes is called the mean degree, denoted by ‹*k*›. It gives a basic measure of how connected the network is and it is calculated using the formula ‹*k*› = 2*M/N*. The factor 2 enters this formula to take into account that every link involves two nodes. For the network in Figure 3, ‹*k*› = 2 11/8 = 2.75.

The examples of *N*, *M* and ‹*k*› refer to global statistics in the sense that they capture simple properties which are averaged over a full network. Different nodes in the network have different degrees and we are also interested in how properties like this are spread over the network, from node to node or from character to character in the present instance. The proportions of nodes that have specific degree values are referred to as the degree distribution. If our network were a regular lattice the links would form a series of squares or rectangles like the left side of Figure 1. Then every interior node would have the same degree. If it were a random network, it would also be quite trivial in that every node has the same probability of taking on a random number of links. Such simple structures are unable to sustain the complex set of interactions that are typical of social or character networks. In a sense, even though random networks appear very different to grids, complex networks are far from both – they are neither random nor regular. In the more realistic settings which manifest them, many nodes have a small number of links, but a few are highly connected and have many links.

We denoted the degree of node A in our simple network of Figure 3 by $k_A$, and we use analogous notation for other nodes. Counting the links of each node we find $k_A = 3$, $k_B = 4$, $k_C = 2$, $k_D = 2$, $k_E = 3$, $k_F = 1$, $k_G = 3$ and $k_H = 4$. We see that one node has degree one ($k=1$), two have degree $k=2$, three have $k=3$ and two have $k=4$. This set of numbers (how the degrees are distributed amongst the nodes) is called the degree distribution. We find it useful to look instead at the complementary cumulative degree distributions $P(k)$. This is the probability that the degree of a node is greater than or equal to the value $k$. In our case, five of the 8 nodes, for example, have degree greater or equal to 3, so we write $P(3) = 5/8$. Doing the same calculation for every degree, we find $P(1) = 8/8 = 1$. $P(2) = 7/8$. $P(4) = 2/8$ and $P(5) = 0$. Plotting these numbers on a graph gives us a visual representation of the spread of connectivity across the network.

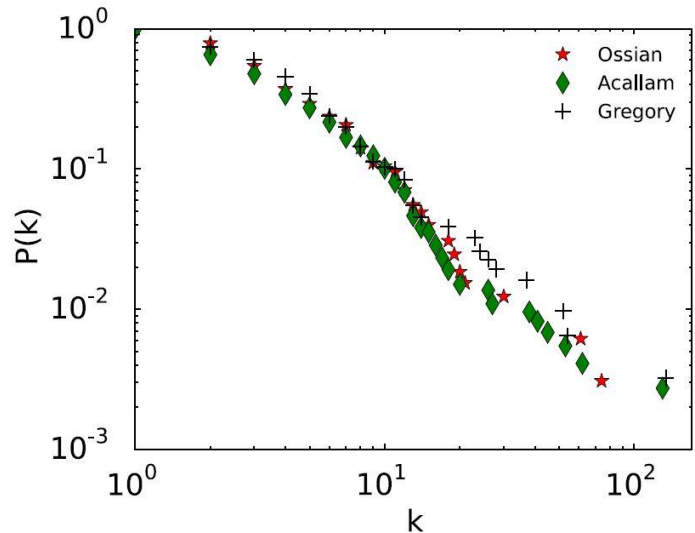

**Figure 4.** The complementary cumulative degree distributions of the full networks indicate that the society depicted in Ossian closely resembles those of the Irish *Acallam na Senórach* and Lady Gregory's text (denoted here by "Gregory").

In a larger network such as that we have generated for *Ossian* and the other narratives in this article, we prefer to present the complementary cumulative degree distributions *P(k)* rather than the degree distribution itself because the former reduces the noisiness in the plot. We present the complementary cumulative degree distributions for *Ossian* and the comparative texts in Figures 4 and 5. These figures represent the main result of this article. In Figure 4 we gather the cumulative degree distributions for *Acallam na Senórach* and Lady Gregory's text alongside that of *Ossian*. The similarities are remarkable. They show that, at least in terms of the degree distributions, the network embedded in *Ossian* is very similar to those of the Irish texts. Moreover, and equally importantly, the network in Lady Gregory's text matches that of *Acallam*. This is as it should be if the former is derived from the latter. Thus we can conclude that (again in terms of the degree distributions) *Ossian* is as similar to *Acallam na Senórach* as is Lady Gregory's text.

We turn our attention to the Classics in Figure 5 where we plot the cumulative degree distribution functions for the Iliad and Odyssey alongside that or Ossian. The dissimilarities are remarkable. Although they appear similar for small values of degree *k*, this is actually meaningless – all cumulative degree distribution functions start at the same place because the degree of all nodes is greater than or equal to zero for any connected network (i.e., all networks start with $P(k) - 1$ for $k = 0$). We can conclude that (again in terms of the degree distributions) Ossian is as dissimilar to Homer's classics.

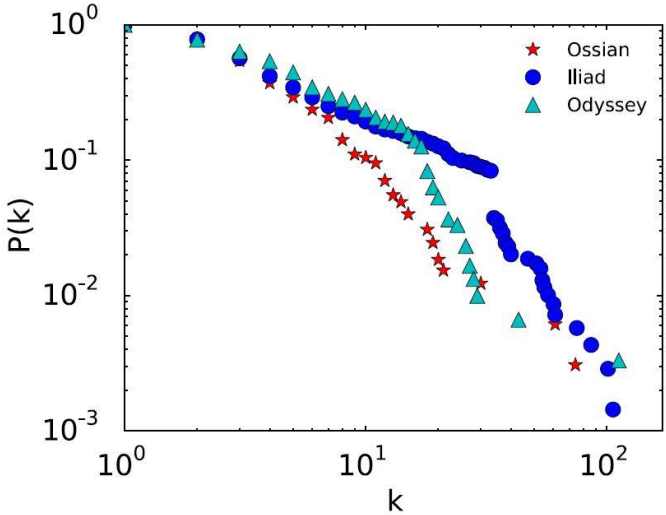

**Figure 5.** The complementary cumulative degree distributions of the full networks indicate that the society depicted in *Ossian* does not resemble those of either the *Iliad* or *Odyssey*.

Visual inspection of these plots delivers the main message of this article: even though Macpherson and his supporters sought to distance Ossian from Irish mythology, the networks unconsciously embedded in the two narratives are remarkably similar. Also, even though Macpherson and his supporters sought to align *Ossian* with Homeric epics, the networks inadvertently embedded in the two narratives are clearly different.

We can pursue a great deal of further analysis to corroborate these conclusions. For example, we can fit standard functions to the various distributions in an attempt to classify them and we can apply the Kolmogorov–Smirnov test to investigate their similarities in a parametric-independent manner. In our 2016 paper,[79] we performed various tests along these lines and they detected clear matches between *Ossian* and both of the Irish texts as well as between *Acallam na Senórach* and Lady Gregory's version. Moreover they detect no significant match between *Ossian* and the Homeric networks.

We can also investigate standard topological measures of the networks underlying each text: statistics which encapsulate a variety of characteristics. We found that Ossianic networks share many universal properties of mythological networks but these measures don't clearly distinguish whether the Ossianic networks are more similar to those of the Irish or the Classics. They do reveal, however, that although *Ossian* is a rather melancholy work, with 666 positive edges (representing friendly interactions) between 309 nodes, and 82 negative edges (representing hostility) between 87 nodes, positive relations and interactions dominate the Ossianic system. Also, when we remove negative links (i.e., remove hostility) from the network it maintains most of its statistical properties, suggesting that although conflict is an important element of the narrative, networks properties are dominated by positive social interactions. For details of these and other statistical properties of the Ossian and how they compare to the Irish and Homeric epics, we again refer the reader to our 2016 paper.

Besides comparing structural properties of the societies underlying the various narratives through degree distributions and network statistics, one may also compare two networks directly using the concept of spectral distances. This concept was originally developed for dynamic biological networks, but it is quite robust and can be considered wherever a quantitative comparison between networks is desired. Again, without going into detail, we discovered that a greater degree of similarity between the Ossianic and Irish graphs was detected than between *Ossian* and the Classics. Therefore, all three approaches—based on parametric fits, Kolmogorov-Smirnov tests, and spectral distances—deliver the same result: the social network structure of *Ossian* bears a measurably closer resemblance to those of the Irish corpus than to the Homeric narratives which Macpherson and his allies strove to parallel.

---

[79] J. Yose, R. Kenna, P. MacCarron, T. Platini, and J. Tonra, "A Networks-Science Investigation into the Epic Poems of Ossian."

## Discussion

In 1738, Daniel Bernoulli developed his kinetic theory of gases, postulating that the macroscopic phenomenon of pressure results from microscopic particles impacting on container walls. This explained Boyle's established law of thermodynamics but was not accepted in the scientific community at the time. Still, it came to form the basis of statistical physics and is taught in introductions to the subject to this day. Daniel was a nephew of Jacob Bernoulli who developed the theory of probability, which in turn was based on games of chance and gambling. In the late 1800s, Ludwig Boltzmann used probability theory to develop modern statistical mechanics, which explains and predicts how the properties of atoms determine the physical properties of matter.[80] However, Boltzmann's theories also met with enormous resistance, from formidable figures such as Max Planck and Ernst Mach. At the time, the tools of probability were still associated with gambling and not trusted.[81]

Jefferson had his own experiments in games of chance: his records document a number of wagers in his younger years, up until

---

[80] L. Boltzmann, "Über die Beziehung zwischen dem zweiten Hauptsatze der mechanischen Wärmetheorie und der Warscheinlichkeitsrechnung respektive den Sätzen über das Wärmegleichgewicht," *Wiener Berichte,* Vol 76, 1877: 373–435.

[81] E.g., E. Johnson, *Anxiety and the Equation: Understanding Boltzmann's Entropy* (Cambridge: *The MIT Press*, 2018).

1785.[82] However, such interests waned in later life. The following comes from part of a series of bills to revise the laws of Virginia in an effort led by Jefferson: "Any person who shall bet or play for money, or other goods, or who shall bet on the hands or sides of those who play at any game in a tavern, racefield, or other place of public resort, shall be deemed an infamous gambler, and shall not be eligible to any office of trust or honor within this state."[83]

Thus, despite later doubts, Jefferson appears not to have been discouraged by moral concerns about probabilistic activities in his early years. It was also in his early years that his main interests in *Ossian* developed and, in this case, the controversies provoked by the work failed to diminish his initial enthusiasm. Amanda Louise Johnson describes him as having "appeared to concede but also sidestep the issue of authenticity."[84] This could be viewed as a signifier of his modernity on this question as it places Jefferson in line with more recent revisionist scholarship on *Ossian*. Indeed, Jefferson's suggestion "that something belated and supposedly manufactured could have the same level of aesthetic value as something that was authentic and original" echoes the development of narrative network studies, where modern interests relate to the comparative complexity of the texts[85] and the manner in which their structures are "crucial to our ability to follow and make sense of a story."[86] As discussed above, we found that the poems of Ossian share many seemingly universal properties of mythological networks.

Our analysis cannot yet clarify any further the question about why *Ossian* appealed to Jefferson. The comparative structural analysis merely reveals a distance between the networks of *Ossian* and the classical texts, but the evidence of Greek and Roman texts in the *Literary Commonplace Book* also shows the enduring influence of Jefferson's classical education. If we cannot argue for the influence of a textual characteristic (network structure) which is unconscious (on an authorial level) and imperceptible (on a reader's level), fuller understanding of the appeal of *Ossian* for Jefferson must be investigated at the levels of text and context. Jefferson appears to have read the characteristic style and diction of *Ossian* as a marker of primitive heroism that influenced his transcription of "Logan's Lament." The contextual situation which views *Ossian* as a response to the domination and subjugation of the Highland character and identity after the Battle of Culloden has more complex and uncertain echoes for Jefferson's interest in the work. Certainly, the narrative of the coloniser and the colonised is present in relations between the Anglo-Americans and the Native Americans, but "Logan's Lament" is transcribed by the dominant, not the dominated, party. The question about what benefit was to be yielded to Jefferson and the Anglo-American republic by valorising Logan's nobility in the Ossianic idiom becomes more complex and uncertain when these rhetorical poles are reversed.

---

[82] https://www.monticello.org/site/research-and-collections/gambling, accessed 1 Jan. 2019.

[83] Thomas Jefferson, "A Bill to Prevent Gaming," *The Papers of Thomas Jefferson* (Princeton: Princeton University Press, 1950), Vol. 2, 306.

[84] Amanda Louise Johnson, "Thomas Jefferson's Ossianic Romance," 20.

[85] P. MacCarron and R. Kenna, *Network Analysis of the Islendinga Sögur: The Sagas of Icelanders*, *European Physical Journal B*, Vol. 86, 2013: 407.

[86] Sandra D. Prado, Silvio R. Dahmen, Ana L.C. Bazzan, Pádraig Mac Carron and Ralph Kenna, "Temporal Network Analysis of Literary Texts," *Advances in Complex Systems*, Vol. 19, 2016: 1650005.

    A purpose of our Ossianic studies thus far was to determine if this modern network-scientific method can deliver something aligned with well-established knowledge in the humanities. We found our various statistical approaches deliver the same outcome; the social network structure embedded in *Ossian* is measurably closer to those of the Irish corpus than to the Homeric narratives, despite attempts by Macpherson and his allies to articulate the opposite. Taking confidence from the soundness demonstrated in the technique, we can move on to the scientific approach of asking new questions based on patterns observed in the data, and appeal to the humanities for the answers. For example, Irish antiquarians readily identified characters such as Fingal with Fionn, Ossian with Oisín and Cuchullin with Cú Chulainn on semantic grounds. The three Ossianic characters in this list are also the three whose betweenness centralities and degrees rank highest in the narrative

network. The question arises to what extent associations between network statistics of lower ranked Ossianic and Fenian characters also correlate with semantic links between them. In other words, ranking characters according network properties might help suggest semantic identifiers as well, thus suggesting possible associations between them across different corpora. However, one would have to ensure not to make the same error made by many bureaucrats who misuse scientometric indicators in the modern academic community; one cannot blindly associate rank with individual attributes such as importance of characters or quality of academic publications.[87] Any parallels between characters in *Ossian* and those in *Acallam na Senórach* (or other narratives) that are suggested by an algorithm would have to be investigated more deeply using traditional humanities approaches, seeking further textual or contextual evidence for how one inspired the other. Indeed, strong positive correlations may be a useful pattern-recognition tool for humanities research while weaker correlations may help educate future managers about the fallacy of blind faith in algorithms. Nonetheless, one should always keep in mind that correlation is not the same as causation.

One wonders what the consequences would have been had Jefferson read *Acallam na Senórach* instead of *Ossian*. At a minimum, by replacing a "fraud" with something more "authentic," it would have removed "a source of embarrassment for the Virginian's biographers who usually dismiss [Jefferson's Ossianism] as youthful enthusiasm"[88] (although it would assuage none of his youthful gambling interests). What would have been the effect of comparing the indigenous peoples to the Irish instead of to the Highlanders, by comparing Logan to the "authentic" Oisín instead of to the "fraudulent" Ossian? The counterfactual idea of Jefferson reading *Acallam* instead of *Ossian* is worth considering somewhat more. With its extant manuscript sources, *Acallamh* provides an authenticity that *Ossian* lacks, but this factor was apparently of little significance for Jefferson. In light of the above, we are left with contemplating the alternative textual and contextual consequences. Jefferson would have read an epic with a similar network structure to *Ossian*, but would he have been similarly enchanted? To answer that, we would need to consider the style and diction of the *Acallamh* and its contexts, and perhaps, the role of Jefferson's Scottish-influenced education as a factor which increased *Ossian*'s appeal. While such textual and contextual analysis appears to be the preserve of the humanities, other interdisciplinary approaches might also be fruitful. In particular, questions about style and diction might be addressed by a partnership of the computational and quantitative methods of stylometry with close reading in order to determine whether *Acallamh* would appeal to Jefferson.

---

[87] R. Kenna, O. Mryglod and B. Berche, "A Scientist's View of Scientometrics: Not Everything that Counts Can be Counted," *Condensed Matter Physics,* Vol. 20, 2017: 1–10.

[88] Amanda Louise Johnson, "Thomas Jefferson's Ossianic Romance," 31.

The interdisciplinary approach applied above requires a combination of numerical and computational skills from the scientific side with the creative and interpretative skills of the humanities. This overcomes the communicative divide as perceived by C.P. Snow in his lecture. The hope that Leavis sought in particular persons can perhaps be more readily found in links between them. The foundations of the third culture envisioned by Snow in his second book may not solely lie in the second part of the term "complexity science" but rather in the more holistic combination of "complexity, curiosity and communication." Indeed, our students, unimpeded, as they are, by traditional disciplinary divides are highly valued and sought after in industry—where the communication skills of humanities coupled with the technical skills of science has given particular advantages.

The idea of using approaches from physics to describe human society emerged in Henri de Saint-Simon's 1803 *Lettres d'un Habitant de Geneve*.[89] His student and collaborator, Auguste Comte, considered as the founder of sociology, famously stated, "Now that the human mind has grasped celestial and terrestrial physics … there remains one science … social physics. This is what men have now most need of." This was echoed over 150 years later when Hawking viewed our time as "the century of complexity." About a century ago, the discipline of physics started to branch into theory and experiment. Whether something similar is happening in the humanities in this, the information age, remains to be seen and whether quantitative approaches of the types described in this article should be categorised as digital humanities or something else is not yet clear. Either way, we have shown how collaborative efforts, with the right kind of communication between two disciplines, can deliver coherent quantitative approaches and analysis to sources from the past. Our experience has further shown how gaps between disciplinary silos and industry can be overcome. These types of applications have enormous potential to help unite the academic disciplines in the pursuit of curiosity-driven inquiries while also forming new links to the corporate world. Looking to the past we may find a way for us all to embrace a bright future.

---

[89] Georg G. Iggers, "Further Remarks about Early Uses of the Term 'Social Science,'" *Journal of the History of Ideas*, Vol. 20, No. 3, 1959: 433–36.